\documentclass[10pt,conference]{IEEEtran}

\usepackage{algorithm}
\usepackage{algpseudocode}
\usepackage{graphicx}
\usepackage{epstopdf}
\usepackage{amsmath}
\usepackage{amssymb,latexsym}
\usepackage{accents}
\usepackage{caption}
\usepackage{subcaption}
\usepackage[framemethod=TikZ]{mdframed}
\usepackage{lipsum}
\usepackage[section]{placeins}
\usepackage[noadjust]{cite}
\usepackage{microtype}
\usepackage{lipsum}
\usepackage{flushend}

\newcommand*\rc{\color[rgb]{0, 0, 0}}
\newcommand*\rca{\color[rgb]{0, 0, 0}}

\IEEEoverridecommandlockouts

\begin{document}

\title{Secrecy Energy Efficiency Optimization for MISO and SISO Communication Networks}
%%%%%%%%%%%%%%%%%%%%%%%%%%%%%%%%%%%%%%%%%%% AUTHER %%%%%%%%%%%%%%%%%%%%%%%%%%%%%%%%%%%%%%%%%%%%%%
\author{
    \IEEEauthorblockN{ Ashkan Kalantari\IEEEauthorrefmark{1}, Sina Maleki\IEEEauthorrefmark{1}, Symeon Chatzinotas\IEEEauthorrefmark{1}, and Bj\"{o}rn Ottersten\IEEEauthorrefmark{1} }
    
		\thanks{This work was supported by the National Research Fund (FNR) of Luxembourg under AFR grant  for the project ``Physical Layer Security in Satellite Communications (ref. 5798109)'' and SeMIGod.} 

    \IEEEauthorblockA{\IEEEauthorrefmark{1} SnT, University of Luxembourg. Emails: \{ashkan.kalantari, sina.maleki, symeon.chatzinotas,  bjorn.ottersten\}@uni.lu}

}

\maketitle
%%%%%%%%%%%%%%%%%%%%%%%%%%%%%%%%%%%%%%%%%%%%% ABSTRACT %%%%%%%%%%%%%%%%%%%%%%%%%%%%%%%%%%%%%%%%%%
\begin{abstract}
Energy-efficiency, high data rates and secure communications are essential requirements of the future wireless networks. In this paper, optimizing the secrecy energy efficiency is considered. The optimal beamformer is designed for a MISO system with and without considering the minimum
required secrecy rate. Further, the optimal power control 
in a SISO system is carried out using an efficient iterative method, and this is followed by analyzing the trade-off between the secrecy energy efficiency and the secrecy rate for both MISO and SISO systems.
\end{abstract}

\begin{IEEEkeywords}
Physical layer security, secrecy rate, secrecy energy efficiency, trade-off, semidefinite programming.
\end{IEEEkeywords}
%
%%%%%%%%%%%%%%%%%%%%%%%%%%%%%%%%%%%%%%%%% Intoduction %%%%%%%%%%%%%%%%%%%%%%%%%%%%%%%%%%%%%%%%%%%%%%%%%
\section{Introduction} \label{Sec:Intro}
Due to the presence of several wireless devices in a specific environment, the transmitted information may be exposed to unintended receivers. Using cryptography in higher layers, a secure transmission can be initiated. Nevertheless, it is probable that a unintended device, which maybe also be a part of the legitimate network, breaks the encryption~\cite{sklavos:Cryptography:2007}. Fortunately, physical layer security techniques can further improve
the security by perfectly securing a transmission rate {\rca using the ``\emph{secrecy rate}'' concept introduced} in~\cite{Wyner:1975}. While security is a concern, power consumption is also another important issue in wireless communications since some wireless devices rely on limited battery power.

There are a wealth of research works in the literature which investigate the energy efficiency in wireless networks such as~\cite{Shuguang:2004,Belmega:2011} and
the references therein. Recently, some research has been done to jointly optimize the secrecy rate and the power consumption. Sum secrecy rate and power are jointly
optimized in~\cite{Ng:2012} to attain a minimum quality of service (QoS).
In~\cite{Xiaoming:2013}, switched beamforming is used to maximize the secrecy outage probability over the consumed power ratio. Powers consumption for a fixed secrecy rate is minimized in~\cite{Wang:2013} for an amplify-and-forward (AF) relay network. The secrecy outage probability over the consumed power is maximized subject to power limit for a large scale AF relay network in~\cite{Jian:2014}. The optimal beamformer for a wiretap channel with multiple-antenna nodes is designed in~\cite{Zhang:2014} to maximize secrecy rate over power ratio.

Here, we consider a multiple-input single-output (MISO) and a single-input single-output (SISO) scenario while a single-antenna unintended receiver, which is part of the network, is listening. The secrecy rate over the power ratio, named ``\emph{secrecy energy efficiency}'' and denoted by $\zeta$, is maximized with and without considering the minimum required secrecy spectral efficiency, denoted by $\eta_0$, at the destination. {\rc  For comparison, we derive the optimal beamformer when zero-forcing (ZF) technique is used to null the signal at the eavesdropper with considering the minimum required secrecy 
spectral efficiency}. Note that the ZF can only be used for the MISO scenario. Furthermore, the trade-off between $\zeta$ and secrecy spectral efficiency, denoted by $\eta$, is studied.

The following issues distinct our work from the most related research. In~\cite{Zhang:2014}, first-order Taylor series expansion and Hadamard inequality are used to approximate the optimal beamfromer for a MIMO system. However, the exact beamformer for the MISO system is derived in this paper. Furthermore, the innermost layer of algorithm in~\cite{Zhang:2014} is based on the singular
value decomposition, and is not applicable to SISO and MISO systems. {\rca In this paper, apart from the MISO system exact beamformer design, exact optimal power allocation for 
the SISO system is also derived.}

The remainder of the paper is organized as follows. In
Section~\ref{Sec:System model}, we introduce the system model. The optimization problems are defined and solved in Sections~\ref{sec:Problem Formulation MISO} and~\ref{sec:Problem Formulation SISO}. In Section~\ref{sec:MISO and SISO trade-off}, the trade-off between secrecy energy efficiency and secrecy spectral efficiency is studied. Numerical results are presented
in Section~\ref{sec:sim}, and the conclusion is drawn in Section~\ref{sec:con}. 
%%%%%%%%%%%%%%%%%%%%%%%%%%%%%%%%%%%%%%%%%%%%%%%%%%%%%%%%%%%%%%%%%%%%%%%%%%%%%%%%%%%%%%%%%%%%%%%%%%
\section{Signal and System Model} \label{Sec:System model}
%%%%%%%%%%%%%%%%%%%%%%%%%%%%%%%%%%%%%%%%%%%%%%%%%%%%%%%%%%%%%%%%%%%%%%%%%%%%%%%%%%%%%%%%%%%%%%%%%%
Consider a wireless communication network comprised of a transmitter denoted by $T$, a receiver denoted by $R$, and an
{\rc unintended} user denoted by $E$. {\rc Note that to obtain the secrecy rate, the legitimate user needs to be aware of the instantaneous channel to the eavesdropper. This knowledge for the most general case with a passive eavesdropper is not practical. In this work, the unintended user is assumed to be part of the network. Therefore, the transmitter $T$ is able to receive the training sequence {\rca from} $E$, in order to estimate {\rca its channel}.}
The signal model and the secrecy rates are derived in the following parts.
%%%%%%%%%%%%%%%%%%%%%%%%%%%%%%%%%%%%%%%%%%%%%%%%%%%%%%%%%%%%%%%%%%%%%%%%%%%%%%%%%%%%%%%%%%%%%%%%%%
\subsection{MISO System}\label{sec:MISO}
%%%%%%%%%%%%%%%%%%%%%%%%%%%%%%%%%%%%%%%%%%%%%%%%%%%%%%%%%%%%%%%%%%%%%%%%%%%%%%%%%%%%%%%%%%%%%%%%%%
Here, we assume that the transmitter employs multiple-antennas. The received signals
at $R$ and $E$ are then as follows
%%%%%%%%%%%%%%%%%%%%%%%%%%%%%%%%%%%%%%%%%%% EQUATION %%%%%%%%%%%%%%%%%%%%%%%%%%%%%%%%%%%%%%%%%%%%%
\begin{align}
{y_R} = {\bf{h}}_{T,R}^T{\bf{w}}x + {n_R},
\label{eqn:D received signal MISO}
\\
{y_E} = {\bf{h}}_{T,E}^T{\bf{w}}x + {n_E},
\label{eqn:E received signal MISO}
\end{align}
%%%%%%%%%%%%%%%%%%%%%%%%%%%%%%%%%%%%%%%%%%%%%%%%%%%%%%%%%%%%%%%%%%%%%%%%%%%%%%%%%%%%%%%%%%%%%%%%%%
where $x$ is the transmitted message, $\bf{w}$ is a vector containing
beamforming gains, ${\bf{h}}_{T,R}$ and ${\bf{h}}_{T,E}$ are
the transmitter's channel gains toward the receiver and eavesdropper, respectively. The additive white Gaussian noise at the receiver and eavesdropper are shown by $n_{R}$
and $n_{E}$, respectively. The random variables $x$, $n_{R}$, and $n_E$ are complex circularly symmetric (c.c.s.) and independent and identically Gaussian distributed (i.i.d.) with~$x\sim\mathcal{CN}(0,1)$, $n_{R}\sim\mathcal{CN}(0, \sigma _{{n_R}}^2 )$, and
$n_{E}\sim\mathcal{CN}(0, \sigma _{{n_E}}^2  )$, respectively, where $\mathcal{CN}$ denotes the complex
normal random variable. The noise powers, $\sigma _{{n_R}}^2$ and $\sigma _{{n_E}}^2$, are equal to $KT_iB$ where $K$ is the boltzman
constant, $T_i$ is the temperature at the corresponding receiver with {\rc $i \in \{R,E\}$}, and $B$ is the transmission bandwidth. Using~\eqref{eqn:D received signal MISO} and~\eqref{eqn:E received signal MISO} and {\rca the result in~\cite{Oggier:2008}}, the secrecy spectral efficiency (or rate in bps/Hz) denoted by $\eta$ is obtained by 
%%%%%%%%%%%%%%%%%%%%%%%%%%%%%%%%%%%%%%%%%%% EQUATION %%%%%%%%%%%%%%%%%%%%%%%%%%%%%%%%%%%%%%%%%%%%%%%
\begin{align}
{\eta _{MISO}} = {\left[ {\log \left( {\frac{{1 + a}}{{1 + b}}} \right)} \right]^ + },
\label{eqn:MISO secrecy rate}
\end{align}
%%%%%%%%%%%%%%%%%%%%%%%%%%%%%%%%%%%%%%%%%%%%%%%%%%%%%%%%%%%%%%%%%%%%%%%%%%%%%%%%%%%%%%%%%%%%%%%%%%
where $a = \frac{{{{\left| {{\bf{h}}_{T,R}^T{\bf{w}}} \right|}^2}}}{{\sigma _{{n_R}}^2}}$, $b = \frac{{{{\left| {{\bf{h}}_{T,E}^T{\bf{w}}} \right|}^2}}}{{\sigma _{{n_E}}^2}}$, and ${\left[ x \right]^ + }$ denotes $\max \left( {x,0} \right)$. In this paper, all the  logarithms are in base two. Further, the operator ${\left[  \cdot  \right]^ + }$ is dropped throughout the paper for the sake of simplicity.
%%%%%%%%%%%%%%%%%%%%%%%%%%%%%%%%%%%%%%%%%%%%%%%%%%%%%%%%%%%%%%%%%%%%%%%%%%%%%%%%%%%%%%%%%%%%%%%%%%
\subsection{SISO System} \label{sec:SISO}
%%%%%%%%%%%%%%%%%%%%%%%%%%%%%%%%%%%%%%%%%%%%%%%%%%%%%%%%%%%%%%%%%%%%%%%%%%%%%%%%%%%%%%%%%%%%%%%%%%
When one antenna is employed at the transmitter, using the result in~\cite{Barros:2006}, the secrecy spectral efficiency, $\eta$, is calculated as
%%%%%%%%%%%%%%%%%%%%%%%%%%%%%%%%%%%%%%%%%%% EQUATION %%%%%%%%%%%%%%%%%%%%%%%%%%%%%%%%%%%%%%%%%%%%%%%
\begin{align}
{\eta _{SISO}} = {\left[ {\log \left( {\frac{{1 + a'}}{{1 + b'}}} \right)} \right]^ + },
\label{eqn:SISO secrecy rate}
\end{align}
%%%%%%%%%%%%%%%%%%%%%%%%%%%%%%%%%%%%%%%%%%%%%%%%%%%%%%%%%%%%%%%%%%%%%%%%%%%%%%%%%%%%%%%%%%%%%%%%%%
where $a' = \frac{{P{{\left| {{h_{T,R}}} \right|}^2}}}{{\sigma _{{n_R}}^2}}$, $b' = \frac{{P{{\left| {{h_{T,E}}} \right|}^2}}}{{\sigma _{{n_E}}^2}}$, $P$ is the transmission power by $T$, and ${h_{{T},{R}}}$ and ${h_{{T},{E}}}$ are the channel gains to the receiver and eavesdropper, respectively.
The statistical characteristics of the message signal and the noise are the same as those in Section~\ref{sec:MISO}.
%%%%%%%%%%%%%%%%%%%%%%%%%%%%%%%%%%%%%%%%%%%%%%%%%%%%%%%%%%%%%%%%%%%%%%%%%%%%%%%%%%%%%%%%%%%%%%%%%%
\section{Problem Formulation: MISO System} \label{sec:Problem Formulation MISO}
%%%%%%%%%%%%%%%%%%%%%%%%%%%%%%%%%%%%%%%%%%%%%%%%%%%%%%%%%%%%%%%%%%%%%%%%%%%%%%%%%%%%%%%%%%%%%%%%%%
In this section, we maximize $\zeta$ in a MISO system by obtaining the optimal beamformer for
the cases with and without QoS constriant at the receiver.
%%%%%%%%%%%%%%%%%%%%%%%%%%%%%%%%%%%%%%%%%%%%%%%%%%%%%%%%%%%%%%%%%%%%%%%%%%%%%%%%%%%%%%%%%%%%%%%%%%
\subsection{With QoS at the Receiver} \label{sec:MISO-QoS}
%%%%%%%%%%%%%%%%%%%%%%%%%%%%%%%%%%%%%%%%%%%%%%%%%%%%%%%%%%%%%%%%%%%%%%%%%%%%%%%%%%%%%%%%%%%%%%%%%%
The metric $\zeta$ is defined as $\eta$ multiplied by bandwidth over the total consumed power ratio as
%%%%%%%%%%%%%%%%%%%%%%%%%%%%%%%%%%%%%%%%%%% EQUATION %%%%%%%%%%%%%%%%%%%%%%%%%%%%%%%%%%%%%%%%%%%%%%%
\begin{align}
\zeta = \frac{{B \eta}}{{ {\left\| {\bf{w}} \right\|^2} + {P_c}}},
\label{eqn:SEE def}
\end{align}
%%%%%%%%%%%%%%%%%%%%%%%%%%%%%%%%%%%%%%%%%%%%%%%%%%%%%%%%%%%%%%%%%%%%%%%%%%%%%%%%%%%%%%%%%%%%%%%%%%
where $P_c$ is the circuit power consumption. We define our problem so as to maximize the secrecy energy 
efficiency subject to the peak power and QoS constraints as follows
%%%%%%%%%%%%%%%%%%%%%%%%%%%%%%%%%%%%%%%%%%% EQUATION %%%%%%%%%%%%%%%%%%%%%%%%%%%%%%%%%%%%%%%%%%%%%%%
\begin{align}
\mathop {\max }\limits_{{\bf{w}}}   \,\,  \zeta  \qquad  \text{s.t.}   \qquad    {\left\| {\bf{w}} \right\|^2} \le {P_{\max }} , \,\,  \eta > {\eta}_0.
\label{eqn:SSE Opt 1}
\end{align}
%%%%%%%%%%%%%%%%%%%%%%%%%%%%%%%%%%%%%%%%%%%%%%%%%%%%%%%%%%%%%%%%%%%%%%%%%%%%%%%%%%%%%%%%%%%%%%%%%%
To design the optimal beamformer, we rewrite~\eqref{eqn:SSE Opt 1} as
%%%%%%%%%%%%%%%%%%%%%%%%%%%%%%%%%%%%%%%%%%% EQUATION %%%%%%%%%%%%%%%%%%%%%%%%%%%%%%%%%%%%%%%%%%%%%%%%
\begin{align}
& \mathop {\max }\limits_{\bf{w}} B\frac{{\log \left( {\frac{{\sigma _{{n_E}}^2}}{{\sigma _{{n_R}}^2}}\frac{{\sigma _{{n_R}}^2 + {{\bf{w}}^H}{\bf{h}}_{T,R}^*{\bf{h}}_{T,R}^T{\bf{w}}}}{{\sigma _{{n_E}}^2 + {{\bf{w}}^H}{\bf{h}}_{T,E}^*{\bf{h}}_{T,E}^T{\bf{w}}}}} \right)}}{{{{\left\| {\bf{w}} \right\|}^2} + {P_c}}}
\nonumber\\
& \,\,  \text{s.t.}   \qquad  \left\| {\bf{w}} \right\|^2 \le {P_{\max }}, \,\,  {{\bf{w}}^H}{\bf{Cw}} \ge  {2^{{\eta}_0}} - 1,
 \label{eqn:MISO opt}
\end{align}
%%%%%%%%%%%%%%%%%%%%%%%%%%%%%%%%%%%%%%%%%%%%%%%%%%%%%%%%%%%%%%%%%%%%%%%%%%%%%%%%%%%%%%%%%%%%%%%%%%
where ${\bf{C}} = \frac{{{\bf{h}}_{T,R}^*{\bf{h}}_{T,R}^T}}{{\sigma _{{n_R}}^2}} - \frac{{{\bf{h}}_{T,E}^*{\bf{h}}_{T,E}^T}}{{\sigma _{{n_E}}^2}}{2^{{\eta}_0}}$.
Using an auxiliary variable as $t=\left\| {\bf{w}} \right\|^2 $,~\eqref{eqn:MISO opt} is reformulated as follows
%%%%%%%%%%%%%%%%%%%%%%%%%%%%%%%%%%%%%%%%%%% EQUATION %%%%%%%%%%%%%%%%%%%%%%%%%%%%%%%%%%%%%%%%%%%%%%%%
\begin{align}
&  \mathop {\max }\limits_{{\bf{w}},0<t \le P_{\max}} {\rm{B}}\frac{{{{\log }}\left( {\frac{{\sigma _{{n_E}}^2}}{{\sigma _{{n_R}}^2}}\frac{{{{\bf{w}}^H}{\bf{Aw}}}}{{{{\bf{w}}^H}{\bf{Bw}}}}} \right)}}{{t + {P_c}}}
\nonumber\\
& \, \text{s.t.}   \,\,  {\left\| {\bf{w}} \right\|^2} = t, \,\,  {{\bf{w}}^H}{\bf{Cw}} \ge {2^{{\eta}_0}} - 1,
 \label{eqn:MISO opt aux}
\end{align}
%%%%%%%%%%%%%%%%%%%%%%%%%%%%%%%%%%%%%%%%%%%%%%%%%%%%%%%%%%%%%%%%%%%%%%%%%%%%%%%%%%%%%%%%%%%%%%%%%%
where ${\bf{A}}={\frac{{\sigma _{{n_R}}^2} }{t}\bf{I} + {\bf{h}}_{T,R}^*{\bf{h}}_{T,R}^T}$ and ${\bf{B}}={\frac{{\sigma _{{n_E}}^2}}{t} \bf{I}+ {\bf{h}}_{T,E}^*{\bf{h}}_{T,E}^T}$. The constraint $\left\| {\bf{w}} \right\|^2 \le {P_{\max }}$ is omitted since the upper limit of the search on variable shall be $P_{\max}$, which satisfy this constraint. To make the last constraint convex,~\eqref{eqn:MISO opt aux} is transformed to a semidefinite programming (SDP) optimization
problem.
%%%%%%%%%%%%%%%%%%%%%%%%%%%%%%%%%%%%%%%%%%% EQUATION %%%%%%%%%%%%%%%%%%%%%%%%%%%%%%%%%%%%%%%%%%%%%%%%
\begin{align}
&\mathop {\max }\limits_{{\bf{W}},0<t \le P_{\max}} {\rm{B}}\frac{{{{\log }}\left( {\frac{{\sigma _{{n_E}}^2}}{{\sigma _{{n_R}}^2}}\frac{{\text{tr}\left( {{\bf{WA}}} \right)}}{{\text{tr}\left( {{\bf{WB}}} \right)}}} \right)}}{{t + {P_c}}}
\nonumber\\
& \, \text{s.t.}   \,\,\,                \text{tr}\left( {\bf{W}} \right) = t, \, \text{tr}\left( {{\bf{WC}}} \right) \ge {2^{{{\eta}_0}}} - 1, \, \bf{W} \succeq 0,
\label{eqn:MISO opt aux SDP}
\end{align}
%%%%%%%%%%%%%%%%%%%%%%%%%%%%%%%%%%%%%%%%%%%%%%%%%%%%%%%%%%%%%%%%%%%%%%%%%%%%%%%%%%%%%%%%%%%%%%%%%%
where $\text{rank}(\bf{W})=1$ constraint is dropped to have a set of convex constraints. Similar to~\cite{Maio:2011}, matrix $\bf{V}$ and scalar $s$ are defined such that ${\bf{V}}=s {\bf{W}}$ and $\text{tr}\left( {s{\bf{WB}}} \right) = 1$. Accordingly,~\eqref{eqn:MISO opt aux SDP} is transformed into
%%%%%%%%%%%%%%%%%%%%%%%%%%%%%%%%%%%%%%%%%%% EQUATION %%%%%%%%%%%%%%%%%%%%%%%%%%%%%%%%%%%%%%%%%%%%%%%%
\begin{align}
& \mathop {\max }\limits_{{\bf{V}},0<t \le P_{\max},s} \frac{{\rm{B}}}{{t + {P_c}}}{\log}\left( {\frac{{\sigma _{{n_E}}^2}}{{\sigma _{{n_R}}^2}}\text{tr}\left( {{\bf{VA}}} \right)} \right)
\nonumber\\
& \, \text{s.t.}   \qquad                                   \text{tr}\left( {\bf{V}} \right) = st, \,\, \text{tr}\left( {{\bf{VC}}} \right) \ge s\left( {{2^{{{\eta}_0}}} - 1} \right),
\nonumber\\
&                               \qquad  \,\,\,\,\,\,\,\,  \text{tr}\left( {{\bf{VB}}} \right) = 1, {\bf{V}} \succeq 0,  s \ge 0.                  
\label{eqn:MISO opt aux SDP sim}
\end{align}
%%%%%%%%%%%%%%%%%%%%%%%%%%%%%%%%%%%%%%%%%%%%%%%%%%%%%%%%%%%%%%%%%%%%%%%%%%%%%%%%%%%%%%%%%%%%%%%%%%
Finally, by considering the auxiliary variable $t$ to be fixed and dropping the $\log$ due to the monotonicity of logarithm function,~\eqref{eqn:MISO opt aux SDP sim} can be solved using SDP along with a one-dimensional search over the variable $t$ where $t \in (0,P_{\max}]$. Since the matrices $\bf{A}$, $\bf{B}$, and $\bf{C}$ in~\eqref{eqn:MISO opt aux SDP sim} are Hermitian positive semidefinite, Theorem 2.{\rca 3} in~\cite{Wenbao} can used to derive an equivalent rank-one solution if the solution to~\eqref{eqn:MISO opt aux SDP sim} satisfies $\text{rank}({\bf{W}}) \ge 3$.

{\rc In order to perform a comparison, we also design the optimal beamforming vector to maximize the secrecy energy 
efficiency when zero-forcing (ZF) strategy is used to null the received signal at the eavesdropper. Using~\eqref{eqn:MISO opt}, the 
{\rca ZF beamformer design} problem can be defined as follows
%%%%%%%%%%%%%%%%%%%%%%%%%%%%%%%%%%%%%%%%%%% EQUATION %%%%%%%%%%%%%%%%%%%%%%%%%%%%%%%%%%%%%%%%%%%%%%%%
\begin{align}
& \mathop {\max }\limits_{\bf{w}} B\frac{{\log \left( {\frac{{\sigma _{{n_R}}^2 + {{\bf{w}}^H}{\bf{h}}_{T,R}^*{\bf{h}}_{T,R}^T{\bf{w}}}}{{\sigma _{{n_R}}^2}}} \right)}}{{{{\left\| {\bf{w}} \right\|}^2} + {P_c}}}
\nonumber\\
& \,  \text{s.t.}  \,\,  \left\| {\bf{w}} \right\|^2 \le {P_{\max }}, \,  {{\bf{w}}^H}{\bf{Cw}} \ge  {2^{{\eta}_0}} - 1, \, {\bf{h}}_{T,E}^T{\bf{w}} = 0.
 \label{eqn:MISO opt ZF 1} 
\end{align}
%%%%%%%%%%%%%%%%%%%%%%%%%%%%%%%%%%%%%%%%%%%%%%%%%%%%%%%%%%%%%%%%%%%%%%%%%%%%%%%%%%%%%%%%%%%%%%%%%%
Using $t = {{\bf{w}}^H}{\bf{w}}$, we get
%%%%%%%%%%%%%%%%%%%%%%%%%%%%%%%%%%%%%%%%%%% EQUATION %%%%%%%%%%%%%%%%%%%%%%%%%%%%%%%%%%%%%%%%%%%%%%%%
\begin{align}
& \mathop {\max }\limits_{\bf{w}} \frac{B}{{t + {P_c}}}\left( {\log \left( {{{\bf{w}}^H}{\bf{Aw}}} \right) - \log \sigma _{{n_R}}^2} \right)
\nonumber\\
& \,  \text{s.t.}  \,\, {\left\| {\bf{w}} \right\|^2} = t,\, {{\bf{w}}^H}{\bf{Cw}} \ge  {2^{{\eta}_0}} - 1, \, {\bf{h}}_{T,E}^T{\bf{w}} = 0,
 \label{eqn:MISO opt ZF 2}
\end{align}
%%%%%%%%%%%%%%%%%%%%%%%%%%%%%%%%%%%%%%%%%%%%%%%%%%%%%%%%%%%%%%%%%%%%%%%%%%%%%%%%%%%%%%%%%%%%%%%%%%
which can be simplified into
%%%%%%%%%%%%%%%%%%%%%%%%%%%%%%%%%%%%%%%%%%% EQUATION %%%%%%%%%%%%%%%%%%%%%%%%%%%%%%%%%%%%%%%%%%%%%%%%
\begin{align}
& \mathop {\max }\limits_{\bf{w}} {{{\bf{w}}^H}{\bf{Aw}}}
\nonumber\\
& \,  \text{s.t.}  \,\, {\left\| {\bf{w}} \right\|^2} = t,\,  {{\bf{w}}^H}{\bf{Cw}} \ge  {2^{{\eta}_0}} - 1, \, {\bf{h}}_{T,E}^T{\bf{w}} = 0.
 \label{eqn:MISO opt ZF 3}
\end{align}
%%%%%%%%%%%%%%%%%%%%%%%%%%%%%%%%%%%%%%%%%%%%%%%%%%%%%%%%%%%%%%%%%%%%%%%%%%%%%%%%%%%%%%%%%%%%%%%%%%
To make the third constraint convex, similar 
to~\eqref{eqn:MISO opt aux},~\eqref{eqn:MISO opt ZF 3} can be transformed into a SDP 
optimization problem as
%%%%%%%%%%%%%%%%%%%%%%%%%%%%%%%%%%%%%%%%%%% EQUATION %%%%%%%%%%%%%%%%%%%%%%%%%%%%%%%%%%%%%%%%%%%%%%%%
\begin{align}
& \mathop {\max }\limits_{\bf{W}} \,\,\,\, \text{tr}\left( {{\bf{WA}}} \right)
\nonumber\\
& \,  \text{s.t.}  \,\, \text{tr}\left( {\bf{W}} \right) = t ,\,  \text{tr}\left( {{\bf{WC}}} \right) \ge  {2^{{\eta}_0}} - 1,
\nonumber\\
&  \,\,\,\,\,\,\,\,\,\, \text{tr}\left( {{\bf{WD}}} \right) = 0,  {\bf{W}} \succeq 0,
 \label{eqn:MISO opt ZF 4}
\end{align}
%%%%%%%%%%%%%%%%%%%%%%%%%%%%%%%%%%%%%%%%%%%%%%%%%%%%%%%%%%%%%%%%%%%%%%%%%%%%%%%%%%%%%%%%%%%%%%%%%%
where ${\bf{D}} = {\bf{h}}_{T,E}^*{\bf{h}}_{T,E}^T$ and the rank-one constraint on $\bf{W}$ is dropped to make the problem convex. Since the 
matrices $\bf{A}$, $\bf{C}$, and $\bf{D}$ in~\eqref{eqn:MISO opt ZF 4} are Hermitian positive semidefinite, Theorem 2.3 in~\cite{Wenbao} can used 
to derive an equivalent rank-one solution if the solution to~\eqref{eqn:MISO opt ZF 4} satisfies $\text{rank}({\bf{W}}) \ge 3$.

If the solution to~\eqref{eqn:MISO opt ZF 4} 
is not rank-one, Theorem 2.3 in~\cite{Wenbao} can be employed to derive an equivalent rank-one 
solution. Problem~\eqref{eqn:MISO opt ZF 4} can be solved using SDP along with a one-dimensional 
search over the variable $t$ where $t \in (0,P_{\max}]$.
}
%%%%%%%%%%%%%%%%%%%%%%%%%%%%%%%%%%%%%%%%%%%%%%%%%%%%%%%%%%%%%%%%%%%%%%%%%%%%%%%%%%%%%%%%%%%%%%%%%%
%%%%%%%%%%%%%%%%%%%%%%%%%%%%%%%%%%%%%%%%%%%%%%%%%%%%%%%%%%%%%%%%%%%%%%%%%%%%%%%%%%%%%%%%%%%%%%%%%%
\subsection{Without QoS at the Receiver} \label{sec:MISO-non-QoS}
%%%%%%%%%%%%%%%%%%%%%%%%%%%%%%%%%%%%%%%%%%%%%%%%%%%%%%%%%%%%%%%%%%%%%%%%%%%%%%%%%%%%%%%%%%%%%%%%%%
Using~\eqref{eqn:MISO opt aux}, the optimal beamformer design problem without considering the QoS is reduced to
%%%%%%%%%%%%%%%%%%%%%%%%%%%%%%%%%%%%%%%%%%% EQUATION %%%%%%%%%%%%%%%%%%%%%%%%%%%%%%%%%%%%%%%%%%%%%%%%
\begin{align}
  \mathop {\max }\limits_{{\bf{w}},0<t \le P_{\max}} {\rm{B}}\frac{{{{\log }}\left( {\frac{{\sigma _{{n_E}}^2}}{{\sigma _{{n_R}}^2}}\frac{{{{\bf{w}}^H}{\bf{Aw}}}}{{{{\bf{w}}^H}{\bf{Bw}}}}} \right)}}{{t + {P_c}}}
 \,\,  \text{s.t.}  \,\, {\left\| {\bf{w}} \right\|^2} = t.
 \label{eqn:MISO opt no QoS}
\end{align}
%%%%%%%%%%%%%%%%%%%%%%%%%%%%%%%%%%%%%%%%%%%%%%%%%%%%%%%%%%%%%%%%%%%%%%%%%%%%%%%%%%%%%%%%%%%%%%%%%%
For a fixed $t$,~\eqref{eqn:MISO opt no QoS} can be written as
%%%%%%%%%%%%%%%%%%%%%%%%%%%%%%%%%%%%%%%%%%% EQUATION %%%%%%%%%%%%%%%%%%%%%%%%%%%%%%%%%%%%%%%%%%%%%%%%
\begin{align}
\mathop {\max }\limits_{\bf{w}} \frac{{\rm{B}}}{{t + {P_c}}}\frac{{\sigma _{{n_E}}^2}}{{\sigma _{{n_R}}^2}}\frac{{{{\bf{w}}^H}{\bf{Aw}}}}{{{{\bf{w}}^H}{\bf{Bw}}}},
 \label{eqn:MISO opt no QoS sim}
\end{align}
%%%%%%%%%%%%%%%%%%%%%%%%%%%%%%%%%%%%%%%%%%%%%%%%%%%%%%%%%%%%%%%%%%%%%%%%%%%%%%%%%%%%%%%%%%%%%%%%%%
where $t \in (0,{P_{\max }}]$. Due to the homogeneity of~\eqref{eqn:MISO opt no QoS}, the constraints on the bamforming vector can be satisfied and thus dropped. The optimal value and the optimal beamforming vector in~\eqref{eqn:MISO opt no QoS sim} are easily derived using
Rayleigh-Ritz~\cite{horn1990matrix} when~\eqref{eqn:MISO opt no QoS sim} is in its standardized form as
%%%%%%%%%%%%%%%%%%%%%%%%%%%%%%%%%%%%%%%%%%% EQUATION %%%%%%%%%%%%%%%%%%%%%%%%%%%%%%%%%%%%%%%%%%%%%%%%
\begin{align}
\mathop {\max }\limits_{\bf{v}} \frac{{\rm{B}}}{{t + {P_c}}}\frac{{\sigma _{{n_E}}^2}}{{\sigma _{{n_R}}^2}}\frac{{{{\bf{v}}^H}{\bf{Dv}}}}{{{{\bf{v}}^H}{\bf{v}}}},
 \label{eqn:MISO opt no QoS sim st}
\end{align}
%%%%%%%%%%%%%%%%%%%%%%%%%%%%%%%%%%%%%%%%%%%%%%%%%%%%%%%%%%%%%%%%%%%%%%%%%%%%%%%%%%%%%%%%%%%%%%%%%%
where ${\bf{v}} = {{\bf{C}}^H}{\bf{w}}$, $\bf{D}={{\bf{C}}^{ - 1}}{\bf{A}}{{\bf{C}}^{ - H}}$, and matrix $\bf{C}$ is the Cholesky decomposition of  matrix ${\bf{B}}$
as ${\bf{B}} = {\bf{C}}{{\bf{C}}^H}$. The optimal beamforming vector is derived as ${{\bf{w}}^ \star } = {{\bf{C}}^{ - H}}{{\bf{v}}^ \star }$ where $\bf{v}^{\star}$ is the eigenvector corresponding to ${{\lambda _{\max }}\left( {{{\bf{C}}^{ - 1}}\bf{A}{{\bf{C}}^{ - H}}} \right)}$. Finally, the
optimal $\zeta$ is obtained in closed-form by
%%%%%%%%%%%%%%%%%%%%%%%%%%%%%%%%%%%%%%%%%%% EQUATION %%%%%%%%%%%%%%%%%%%%%%%%%%%%%%%%%%%%%%%%%%%%%%%%
\begin{align}
\zeta{^ \star } =  B \frac{{{{\log }}\left( {\frac{{\sigma _{{n_E}}^2}}{{\sigma _{{n_R}}^2}}{\lambda _{\max }}\left( {{{\bf{C}}^{ - 1}}{\bf{A}}{{\bf{C}}^{ - {\bf{H}}}}} \right)} \right)}}{{t + {P_c}}}.
 \label{eqn:zeta opt no QoS}
\end{align}
%%%%%%%%%%%%%%%%%%%%%%%%%%%%%%%%%%%%%%%%%%%%%%%%%%%%%%%%%%%%%%%%%%%%%%%%%%%%%%%%%%%%%%%%%%%%%%%%%%
{\rca Employing} a one-dimensional search over $t \in (0,{P_{\max }}]$ and using~\eqref{eqn:zeta opt no QoS}, the optimal value of~\eqref{eqn:MISO opt no QoS sim st} is found.
%%%%%%%%%%%%%%%%%%%%%%%%%%%%%%%%%%%%%%%%%%%%%%%%%%%%%%%%%%%%%%%%%%%%%%%%%%%%%%%%%%%%%%%%%%%%%%%%%%
\section{Problem Formulation: SISO System} \label{sec:Problem Formulation SISO}
%%%%%%%%%%%%%%%%%%%%%%%%%%%%%%%%%%%%%%%%%%%%%%%%%%%%%%%%%%%%%%%%%%%%%%%%%%%%%%%%%%%%%%%%%%%%%%%%%%
In the SISO case, the beamformer design is reduced to scalar power control. Similar to~\eqref{eqn:SSE Opt 1}, the optimization problem for SISO system is
defined as
%%%%%%%%%%%%%%%%%%%%%%%%%%%%%%%%%%%%%%%%%%% EQUATION %%%%%%%%%%%%%%%%%%%%%%%%%%%%%%%%%%%%%%%%%%%%%
\begin{align}
\mathop {\max }\limits_P    \,\,\,\,   B\frac{{\log \left( {\frac{{\sigma _{{n_E}}^2}}{{\sigma _{{n_R}}^2}}\frac{{\sigma _{{n_R}}^2 + P{{\left| {{h_{T,R}}} \right|}^2}}}{{\sigma _{{n_E}}^2 + P{{\left| {{h_{T,E}}} \right|}^2}}}} \right)}}{{{P_c} + P}}
 \,\, \text{s.t.}   \,\,  P_{\min}  \leq P \leq {P_{\max }},
\label{eqn:SSE Opt 3}
\end{align}
%%%%%%%%%%%%%%%%%%%%%%%%%%%%%%%%%%%%%%%%%%%%%%%%%%%%%%%%%%%%%%%%%%%%%%%%%%%%%%%%%%%%%%%%%%%%%%%%%%
where ${P_{\min }} = \frac{{{2^{{\eta _0}}} - 1}}{\alpha }$ is obtained from the minimum QoS constraint, and it is assumed that $\alpha = \frac{{{{\left| {{h_{T,R}}} \right|}^2}}}{{\sigma _{{n_R}}^2}} - \frac{{{{\left| {{h_{T,E}}} \right|}^2}}}{{\sigma _{{n_E}}^2}}{2^{{{\eta}_0}}} > 0$. The numerator in the objective of~\eqref{eqn:SSE Opt 3} is concave since the argument of the logarithm is concave for $P \ge 0$ and $\frac{{{{\left| {{h_{T,R}}} \right|}^2}}}{{\sigma _{{n_R}}^2}} > \frac{{{{\left| {{h_{T,E}}} \right|}^2}}}{{\sigma _{{n_E}}^2}}$, which are granted in our problem, and the denumerator is affine. Hence,~\eqref{eqn:SSE Opt 3} is categorized as a family member of fractional programming problems known as~``\emph{concave fractional program}'' where a local optimum is a global one~\cite{Siegfried:1983}. Here, we solve~\eqref{eqn:SSE Opt 3} using an iterative (parametric) algorithm named Dinkelbach~\cite{Dinkelbach:Dinkelbach}. For the sake of simplicity, we mention the values related to ${\left| {{h_{T,R}}} \right|^2}$ and ${\left| {{h_{T,E}}} \right|^2}$ by $a$ and $b$, respectively.
%%%%%%%%%%%%%%%%%%%%%%%%%%%%%%%%%%%%%%%%%%%% EQUATION %%%%%%%%%%%%%%%%%%%%%%%%%%%%%%%%%%%%%%%%%%%%%
%%%%%%%%%%%%%%%%%%%%%%%%%%%%%%%%%%%%%%%%%%%%%%%%%%%%%%%%%%%%%%%%%%%%%%%%%%%%%%%%%%%%%%%%%%%%%%%%%%%
%Consequently, the optimal value of $x$ is calculated using an iterative approach.
According to~\cite{Dinkelbach:Dinkelbach}, after dropping the constant $B$,~\eqref{eqn:SSE Opt 3} is written as
%%%%%%%%%%%%%%%%%%%%%%%%%%%%%%%%%%%%%%%%%%% EQUATION %%%%%%%%%%%%%%%%%%%%%%%%%%%%%%%%%%%%%%%%%%%%%
\begin{align}
& F\left( q \right) = \mathop {\max }\limits_{P \in S} \log \left( {\frac{{\sigma _{{n_E}}^2}}{{\sigma _{{n_R}}^2}}\frac{{\sigma _{{n_R}}^2 + Pa}}{{\sigma _{{n_E}}^2 + Pb}}} \right) - q\left( {{P_c} + P} \right),
\label{eqn:linear equation}
\\
& q = \frac{{f\left( P \right)}}{{g\left( P \right)}},
\label{eqn:q def}
\end{align}
%%%%%%%%%%%%%%%%%%%%%%%%%%%%%%%%%%%%%%%%%%%%%%%%%%%%%%%%%%%%%%%%%%%%%%%%%%%%%%%%%%%%%%%%%%%%%%%%%%
where $f(P)$ and $g(P)$ are the numerator and denumerator of~\eqref{eqn:SSE Opt 3}, respectively. Also, $S$ shows the feasible domain of $P$. To
calculate the optimal $P$ for~\eqref{eqn:linear equation}, denoted by $P^{\star}$, the derivative of $F(q)$ with respect to $P$ is calculated as follows
%%%%%%%%%%%%%%%%%%%%%%%%%%%%%%%%%%%%%%%%%%% EQUATION %%%%%%%%%%%%%%%%%%%%%%%%%%%%%%%%%%%%%%%%%%%%%%%
\begin{align}
 \frac{{\partial F}}{{\partial P}} = & - abq\beta {P^2} + Pq\beta \left( { - a\sigma _{{n_E}}^2 - b\sigma _{{n_R}}^2} \right)
\nonumber\\
& + a\sigma _{{n_E}}^2 - q\beta \sigma _{{n_R}}^2\sigma _{{n_E}}^2 - b\sigma _{{n_R}}^2,
\label{eqn:der}
\end{align}
%%%%%%%%%%%%%%%%%%%%%%%%%%%%%%%%%%%%%%%%%%%%%%%%%%%%%%%%%%%%%%%%%%%%%%%%%%%%%%%%%%%%%%%%%%%%%%%%%%
which is a quadratic equation with a closed-form solution as
%%%%%%%%%%%%%%%%%%%%%%%%%%%%%%%%%%%%%%%%%%% EQUATION %%%%%%%%%%%%%%%%%%%%%%%%%%%%%%%%%%%%%%%%%%%%%%%%
\begin{align}
& {P_{1,2}} = \frac{{q\left( {a\sigma _{{n_E}}^2 + b\sigma _{{n_R}}^2} \right) \pm \sqrt \Delta  }}{{ - 2abq}}, \,\, \beta=\text{Ln}2,
\nonumber\\
&\Delta  = {q^2}{\left( {a\sigma _{{n_E}}^2 + b\sigma _{{n_R}}^2} \right)^2} + 4abq\left( {a\sigma _{{n_E}}^2 - q\sigma _{{n_R}}^2\sigma _{{n_E}}^2 - b\sigma _{{n_R}}^2} \right).
\label{eqn:closed_form P}
\end{align}
%%%%%%%%%%%%%%%%%%%%%%%%%%%%%%%%%%%%%%%%%%%%%%%%%%%%%%%%%%%%%%%%%%%%%%%%%%%%%%%%%%%%%%%%%%%%%%%%%%
Since $P_1$ in~\eqref{eqn:closed_form P} is always negative, ${P^{\star}}$ is derived as
%%%%%%%%%%%%%%%%%%%%%%%%%%%%%%%%%%%%%%%%%%% EQUATION %%%%%%%%%%%%%%%%%%%%%%%%%%%%%%%%%%%%%%%%%%%%%
\begin{align}
P^ \star  = \left\{ {\begin{array}{*{20}{c}}
P_2            &       P_2 \in S,
\\
\mathop {\arg }\limits_P \mathop {\max }\limits_{P \in \left\{ {{P_{\min }},{P_{\max }}} \right\}} \,\,F\left( q \right)  &  P_2  \notin  S,
\end{array}} \right.
\label{P Opt case 1 & 2}
\end{align}
%%%%%%%%%%%%%%%%%%%%%%%%%%%%%%%%%%%%%%%%%%%%%%%%%%%%%%%%%%%%%%%%%%%%%%%%%%%%%%%%%%%%%%%%%%%%%%%%%%
where ${P_2} = \frac{{q\left( {a\sigma _{{n_E}}^2 + b\sigma _{{n_R}}^2} \right) - \sqrt \Delta  }}{{ - 2abq}}$. The procedure to
solve~\eqref{eqn:SSE Opt 3} using Dinkelbach method is summarized in Algorithm~\ref{Alg Par}.
%%%%%%%%%%%%%%%%%%%%%%%%%%%%%%%%%%%%%%%%%%% ALGORITHM %%%%%%%%%%%%%%%%%%%%%%%%%%%%%%%%%%%%%%%%%%%%
\algnewcommand{\algorithmicgoto}{\textbf{Go to}}%
\algnewcommand{\Goto}[1]{\algorithmicgoto~\ref{#1}}%
\begin{algorithm}[t]
\caption{Iterative approach to solve~\eqref{eqn:SSE Opt 3}}
\begin{algorithmic}[1]
\State Initialize $n=0$;
\State Pick any $P_n \in S $;
\State Derive $q_n$ using~\eqref{eqn:q def}; \label{alg:q}
\State Derive $P^{\star}_n$ using~\eqref{P Opt case 1 & 2} and calculate $F(q_n)$ using~\eqref{eqn:linear equation}; \label{alg:p}
\If{ $F(q_n)\geq \delta$ }
\State $n=n+1$;
\State  \Goto{alg:q};
\EndIf
%\State end. \label{end};
\end{algorithmic}
\label{Alg Par}
\end{algorithm}
%%%%%%%%%%%%%%%%%%%%%%%%%%%%%%%%%%%%%%%%%%%%%%%%%%%%%%%%%%%%%%%%%%%%%%%%%%%%%%%%%%%%%%%%%%%%%%%%%%
Using the closed-form solution of~\eqref{eqn:linear equation} {\rca given in~\eqref{P Opt case 1 & 2}}, the following recursive relation is used to merge Steps~\ref{alg:q} and~\ref{alg:p} of Algorithm~\ref{Alg Par} as
%%%%%%%%%%%%%%%%%%%%%%%%%%%%%%%%%%%%%%%%%%% EQUATION %%%%%%%%%%%%%%%%%%%%%%%%%%%%%%%%%%%%%%%%%%%%%%%%
\begin{align}
{P_{n + 1}} =\frac{{\frac{{f\left( {{P_n}} \right)}}{{g\left( {{P_n}} \right)}}\left( {a\sigma _{{n_E}}^2 + b\sigma _{{n_R}}^2} \right) - \sqrt {{\Delta _n}} }}{{ - 2ab\frac{{f\left( {{P_n}} \right)}}{{g\left( {{P_n}} \right)}}}}.
 \label{eqn:iterative}
\end{align}
%%%%%%%%%%%%%%%%%%%%%%%%%%%%%%%%%%%%%%%%%%%%%%%%%%%%%%%%%%%%%%%%%%%%%%%%%%%%%%%%%%%%%%%%%%%%%%%%%%
It is proven in~\cite{Dinkelbach:Dinkelbach} that Algorithm~\ref{Alg Par} converges. In addition, since a local optimum for a concave fractional program is the global optimum, and~\eqref{eqn:SSE Opt 3} falls into this category, the solution found using Algorithm~\ref{Alg Par} is a global optimum.
%%%%%%%%%%%%%%%%%%%%%%%%%%%%%%%%%%%%%%%%%%%%%%%%%%%%%%%%%%%%%%%%%%%%%%%%%%%%%%%%%%%%%%%%%%%%%%%%%%
%%%%%%%%%%%%%%%%%%%%%%%%%%%%%%%%%%%%%%%%%%%%%%%%%%%%%%%%%%%%%%%%%%%%%%%%%%%%%%%%%%%%%%%%%%%%%%%%%%
\section{Trade-off between $\zeta$ and $\eta$} \label{sec:MISO and SISO trade-off}
%%%%%%%%%%%%%%%%%%%%%%%%%%%%%%%%%%%%%%%%%%%%%%%%%%%%%%%%%%%%%%%%%%%%%%%%%%%%%%%%%%%%%%%%%%%%%%%%%%
In this section, we study the trade-off between secrecy energy efficiency and secrecy spectral efficiency (i.e. $\zeta$ and $\eta$) for MISO and SISO systems.
%%%%%%%%%%%%%%%%%%%%%%%%%%%%%%%%%%%%%%%%%%%%%%%%%%%%%%%%%%%%%%%%%%%%%%%%%%%%%%%%%%%%%%%%%%%%%%%%%%
\subsection{MISO System}  \label{subsec:relating SEE and SSE:MISO}
%%%%%%%%%%%%%%%%%%%%%%%%%%%%%%%%%%%%%%%%%%%%%%%%%%%%%%%%%%%%%%%%%%%%%%%%%%%%%%%%%%%%%%%%%%%%%%%%%%
To find the trade-off between $\zeta$ and $\eta$, we solve the optimal beamforming design problem to maximize
$\zeta$ and $\eta$ separately for a specific power constraint, $P$. As a result,
the pair $(\zeta , \eta)$ is available for different values of $P$. For $\zeta$, the optimization problem is as follows
%%%%%%%%%%%%%%%%%%%%%%%%%%%%%%%%%%%%%%%%%%% EQUATION %%%%%%%%%%%%%%%%%%%%%%%%%%%%%%%%%%%%%%%%%%%%%%%%
\begin{align}
 \mathop {\max }\limits_{\bf{w}} {\mkern 1mu} {\mkern 1mu}   B\frac{{{{\log }_2}\left( {\frac{{\sigma _{{n_E}}^2}}{{\sigma _{{n_R}}^2}}\frac{{\sigma _{{n_R}}^2 + {{\bf{w}}^H}{\bf{h}}_{T,R}^*{\bf{h}}_{T,R}^T{\bf{w}}}}{{\sigma _{{n_E}}^2 + {{\bf{w}}^H}{\bf{h}}_{T,E}^*{\bf{h}}_{T,E}^T{\bf{w}}}}} \right)}}{{{P} + {P_c}}}
 \,\, \text{s.t.}   \,\,  \left\| {\bf{w}} \right\|^2 = P.
 \label{eqn:MISO TO 2}
\end{align}
%%%%%%%%%%%%%%%%%%%%%%%%%%%%%%%%%%%%%%%%%%%%%%%%%%%%%%%%%%%%%%%%%%%%%%%%%%%%%%%%%%%%%%%%%%%%%%%%%%
Using the constraint in~\eqref{eqn:MISO TO 2}, we conclude that $\frac{{{{\bf{w}}^H}{\bf{w}}}}{{{P}}} = 1$ which helps us homogenize~\eqref{eqn:MISO TO 2} as
%%%%%%%%%%%%%%%%%%%%%%%%%%%%%%%%%%%%%%%%%%% EQUATION %%%%%%%%%%%%%%%%%%%%%%%%%%%%%%%%%%%%%%%%%%%%%%%%
\begin{align}
 \mathop {\max }\limits_{\bf{w}} {\rm{B}}\frac{{{{\log }_2}\left( {\frac{{\sigma _{{n_E}}^2}}{{\sigma _{{n_R}}^2}}\frac{{{{\bf{w}}^H}{\bf{Aw}}}}{{{{\bf{w}}^H}{\bf{Bw}}}}} \right)}}{{P + {P_c}}}
 \,\, \text{s.t.}   \,\,  \left\| {\bf{w}} \right\|^2 = P,
 \label{eqn:MISO TO 3}
\end{align}
%%%%%%%%%%%%%%%%%%%%%%%%%%%%%%%%%%%%%%%%%%%%%%%%%%%%%%%%%%%%%%%%%%%%%%%%%%%%%%%%%%%%%%%%%%%%%%%%%%
where, ${\bf{A}} = \frac{{\sigma _{{n_R}}^2}}{P} {\bf{I}}+ {\bf{h}}_{T,R}^*{\bf{h}}_{T,R}^T$ and
${\bf{B}} = \frac{{\sigma _{{n_E}}^2}}{P} {\bf{I}} + {\bf{h}}_{T,E}^*{\bf{h}}_{T,E}^T$. Similar to~\eqref{eqn:MISO opt no QoS}, the $\log$ and the power constraint
can be dropped. Similar to the solution to~\eqref{eqn:MISO opt no QoS sim st}, the optimal beamforming vector shall be ${{\bf{w}}^ \star } = {{\bf{C}}^{ - H}}{{\bf{v}}^ \star }$ where $\bf{v}^{\star}$ is the eigenvector corresponding to ${{\lambda _{\max }}\left( {{{\bf{C}}^{ - 1}}\bf{A}{{\bf{C}}^{ - H}}} \right)}$. The final closed-form solution for $\zeta{^ \star }$ is
%%%%%%%%%%%%%%%%%%%%%%%%%%%%%%%%%%%%%%%%%%% EQUATION %%%%%%%%%%%%%%%%%%%%%%%%%%%%%%%%%%%%%%%%%%%%%%%%
\begin{align}
\zeta{^ \star } =  B \frac{{{{\log }}\left( {\frac{{\sigma _{{n_E}}^2}}{{\sigma _{{n_R}}^2}}{\lambda _{\max }}\left( {{{\bf{C}}^{ - 1}}{\bf{A}}{{\bf{C}}^{ - {\bf{H}}}}} \right)} \right)}}{{P + {P_c}}}.
 \label{eqn:CF trade-off}
\end{align}
%%%%%%%%%%%%%%%%%%%%%%%%%%%%%%%%%%%%%%%%%%%%%%%%%%%%%%%%%%%%%%%%%%%%%%%%%%%%%%%%%%%%%%%%%%%%%%%%%%
The optimal beamforming vector for $\eta{^ \star }$ shall be the same as for $\zeta{^ \star }$ and 
the optimal value of $\eta$ can be derived similar to the one for $\zeta$. Hence, the 
pair $(\zeta , \eta)$ is available.
%%%%%%%%%%%%%%%%%%%%%%%%%%%%%%%%%%%%%%%%%%%%%%%%%%%%%%%%%%%%%%%%%%%%%%%%%%%%%%%%%%%%%%%%%%%%%%%%%%
\subsection{SISO System}  \label{subsec:relating SEE and SSE:SISO}
%%%%%%%%%%%%%%%%%%%%%%%%%%%%%%%%%%%%%%%%%%%%%%%%%%%%%%%%%%%%%%%%%%%%%%%%%%%%%%%%%%%%%%%%%%%%%%%%%%
%%%%%%%%%%%%%%%%%%%%%%%%%%%%%%%%%%%%%%%%%%%%%%%%%%%%%%%%%%%%%%%%%%%%%%%%%%%%%%%%%%%%%%%%%%%%%%%%%%
By deriving $P$ with respect to $\eta$ using~\eqref{eqn:SISO secrecy rate} as $P = \frac{{\sigma _{{n_R}}^2\sigma _{{n_E}}^2\left( {{2^{\eta}} - 1} \right)}}{{\sigma _{{n_E}}^2a - \sigma _{{n_R}}^2b{2^{\eta}}}}$,
the relation between $\zeta$ and $\eta$ is calculated using~\eqref{eqn:SEE def} as follows
%%%%%%%%%%%%%%%%%%%%%%%%%%%%%%%%%%%%%%%%%%% EQUATION %%%%%%%%%%%%%%%%%%%%%%%%%%%%%%%%%%%%%%%%%%%%%%%%
\begin{align}
\zeta = \frac{{B\eta\left( {\sigma _{{n_E}}^2a - \sigma _{{n_R}}^2b{2^{\eta}}} \right)}}{{\sigma _{{n_R}}^2\sigma _{{n_E}}^2\left( {{2^{\eta}} - 1} \right) + {P_c}\left( {\sigma _{{n_E}}^2a - \sigma _{{n_R}}^2b{2^{\eta}}} \right)}}.
 \label{eqn:SEE wrt SSE}
\end{align}
%%%%%%%%%%%%%%%%%%%%%%%%%%%%%%%%%%%%%%%%%%%%%%%%%%%%%%%%%%%%%%%%%%%%%%%%%%%%%%%%%%%%%%%%%%%%%%%%%%%
By solving $\frac{{d\zeta }}{{d\eta }}=0$ using numerical methods, $\eta$ corresponding to the optimal $\zeta$ can be derived.
%%%%%%%%%%%%%%%%%%%%%%%%%%%%%%%%%%%%%%%%%%%%%%%%%%%%%%%%%%%%%%%%%%%%%%%%%%%%%%%%%%%%%%%%%%%%%%%%%%%
%\vspace{-3cm}
%%%%%%%%%%%%%%%%%%%%%%%%%%%%%%%%%%%%%%%%%%% Link Budgect Table %%%%%%%%%%%%%%%%%%%%%%%%%%%%%%%%%%%
%\begin{table}[]
%\caption{System and link parameters}            % title of Table
%\centering                                                % used for centering table
%{\small \begin{tabular}{|l| l|}                                  % centered columns (2 columns)
%\hline
%Parameter & Value  \\ [0.5ex] % inserts table
%\hline                                                          % inserts single horizontal line
%\hline
%Channel model ($\mathcal{CN}(0,1)$)      &  Quasi-static block fading \\
%\hline
%Path loss (dB)      &  $128.1 + 37.6 \, \log_{10}d$ \\
%\hline
%Bandwidth (MHz), B  & 20  \\
%\hline
%Circuit power, (unit of power), $P_c$      &  5  \\
%\hline
%Maximum power, $P_{\max}$      &  50  \\
%\hline
%Receiver temperature in (Kelvin), T            &   298   \\
%\hline
%Tolerance error for Dinkelbach, $\delta$             &   $10^{-3}$   \\
%\hline
%\end{tabular}}
%\label{table:Link budget parameters}
%\end{table}
%%%%%%%%%%%%%%%%%%%%%%%%%%%%%%%%%%%%%%%%%%%%%%%%%%%%%%%%%%%%%%%%%%%%%%%%%%%%%%%%%%%%%%%%%%%%%%%%%%%%
%%%%%%%%%%%%%%%%%%%%%%%%%%%%%%%%%%%%Simulations%%%%%%%%%%%%%%%%%%%%%%%%%%%%%%%%%%%%%%%%%%%%%%%%%%%%%%%%
\section{Simulation Results}  \label{sec:sim}
%%%%%%%%%%%%%%%%%%%%%%%%%%%%%%%%%%%%%%%%%%%%%%%%%%%%%%%%%%%%%%%%%%%%%%%%%%%%%%%%%%%%%%%%%%%%%%%%%%%
In this section, we present numerical examples to investigate the secrecy energy
efficiency and its trade-off with the secrecy spectral efficiency. The simulations' parameters are as follows. Distance from the transmitter to receiver and eavesdropper, $d$, is considered to 
be $2$ km, Quasi-static block fading channel model as $\mathcal{CN}(0,1)$, path loss is $128.1 + 37.6\log_{10}d$ dB~\cite{3GPP}, bandwidth is $20$ MHz, $P_c=5$, $P_{\max}=50$, receiver noise temperature 
is $298$ K, tolerance error for Dinkelbach algorithm is $\delta=10^{-3}$, and $N$ is the number of antennas. 
%The rest of the parameters are presented in Table~\ref{table:Link budget parameters}~\cite{3GPP}. 
If the secrecy rate is negative, it is considered to be zero. For the figures presenting the average graphs, 
enough amount of channels are generated and the average of the resultant metrics are considered.
%%%%%%%%%%%%%%%%%%%%%%%%%%%%%%%%%%%%%%%%%%%%%%% FIGURE  %%%%%%%%%%%%%%%%%%%%%%%%%%%%%%%%%%%%%%%%%%%%%%%
\begin{figure}[]
  \centering
  \includegraphics[width=9cm]{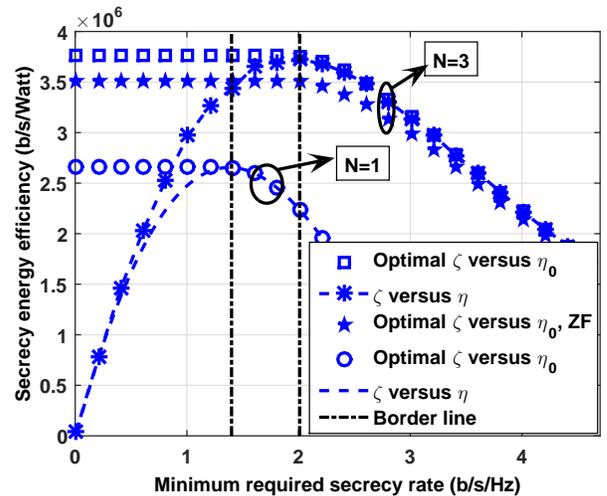}
  \caption{Optimal $\zeta$ versus $\eta_0$ and $\zeta$ versus $\eta$ graphs.}
  \label{fig:SEE vs SSE trade-off SISO MISO}
\end{figure}
%%%%%%%%%%%%%%%%%%%%%%%%%%%%%%%%%%%%%%%%%%%%%%%%%%%%%%%%%%%%%%%%%%%%%%%%%%%%%%%%%%%%%%%%%%%%%%%%%%%%
%%%%%%%%%%%%%%%%%%%%%%%%%%%%%%%%%%%%%%%%%%%%%%% FIGURE  %%%%%%%%%%%%%%%%%%%%%%%%%%%%%%%%%%%%%%%%%%%%%%%
\begin{figure}[]
  \centering
  \includegraphics[width=9cm]{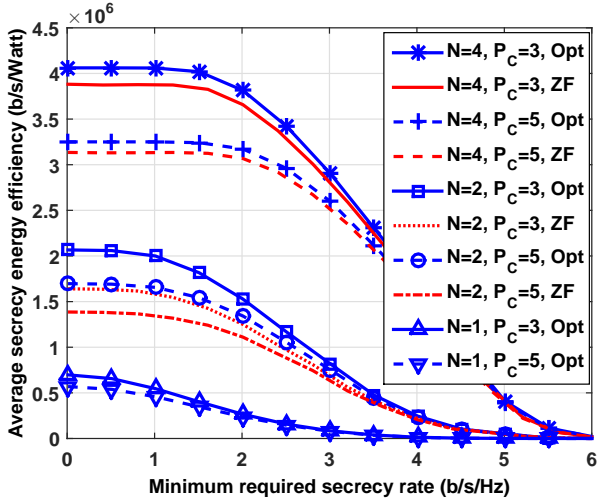}
  \caption{Average $\zeta$ versus ${\eta}_0$ for different $N$ and $P_c$.}
  \label{fig:SEE vs R0 PC}
\end{figure}
%%%%%%%%%%%%%%%%%%%%%%%%%%%%%%%%%%%%%%%%%%%%%%%%%%%%%%%%%%%%%%%%%%%%%%%%%%%%%%%%%%%%%%%%%%%%%%%%%%%%
In the first simulation scenario, the secrecy energy efficiency and secrecy spectral efficiency trade-off is studied. Optimal $\zeta$ versus the minimum required
$\eta$ graphs as well as the graphs related to the trade-off between $\zeta$ and $\eta$ are presented in Fig.~\ref{fig:SEE vs SSE trade-off SISO MISO} using a single channel realization. Two different regions are defined in Fig.~\ref{fig:SEE vs SSE trade-off SISO MISO} using a border line. The border line defines the optimal operating point in terms of $\zeta$. In the left-hand side region, increasing $\eta$ also increases $\zeta$. Hence, to get a higher $\zeta$, the
secrecy rate can be increased, which is desirable. However, the mechanism between $\zeta$ and $\eta$ changes in the right-hand side of
Fig.~\ref{fig:SEE vs SSE trade-off SISO MISO}. After the optimal point of $\zeta$, increasing $\eta$ demands more power which is higher than the optimal
power value for $\zeta$. Therefore, as $\eta$ increases, $\zeta$ falls below the optimal value which is opposite to the procedure in the left-hand side, and the trade-off is clear. {\rc Also, it is observed that ZF results in a lower secrecy energy efficiency. Nevertheless, as the minimum required secrecy rate increases, the performance of the ZF approaches the 
primary scheme{\rca, i.e., optimal beamformer design.}
%%%%%%%%%%%%%%%%%%%%%%%%%%%%%%%%%%%%%%%%%%%%%%% FIGURE  %%%%%%%%%%%%%%%%%%%%%%%%%%%%%%%%%%%%%%%%%%%%%%%
\begin{figure}[]
  \centering
  \includegraphics[width=9cm]{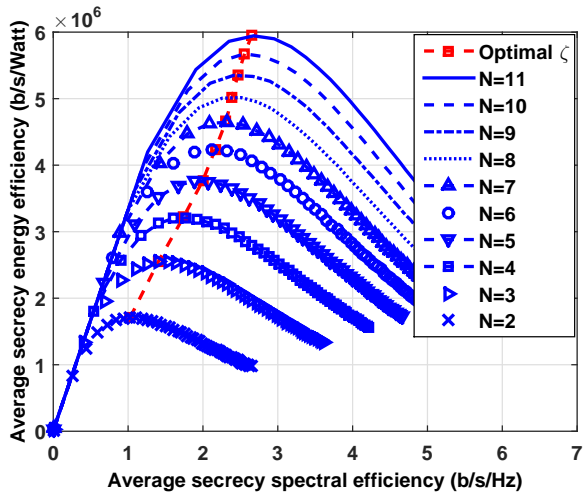}
  \caption{$\zeta$ and $\eta$ relation for different antennas.}
  \label{fig:trade-off MISO}
\end{figure}
%%%%%%%%%%%%%%%%%%%%%%%%%%%%%%%%%%%%%%%%%%%%%%%%%%%%%%%%%%%%%%%%%%%%%%%%%%%%%%%%%%%%%%%%%%%%%%%%%%%%%%

For the second scenario, average $\zeta$ versus the minimum required $\eta$ is investigated for different numbers of antennas, and circuit powers. The related graphs are depicted in Fig.~\ref{fig:SEE vs R0 PC}. As it is shown, increasing the number of antennas results in increasing the optimal value of $\zeta$  and makes it stable for a longer range of $\eta_0$. Further, we can see that decreasing $P_{c}$ leads to higher secrecy energy efficiency, and this is more significant for higher number of antennas. {\rc Similar to the result in Fig.~\ref{fig:SEE vs SSE trade-off SISO MISO}, ZF scheme shows a sub-optimal performance. ZF's performance gets closer to the optimal scheme as the circuit power, $P_c$, increases. Interestingly, for fewer number of antennas, the gap between the performance of the ZF and the optimal scheme even gets larger. {\rca This is due to less degrees of freedom for the ZF beamformer design as the number of antennas decreases.}}
To investigate the trade-off between $\zeta$ and $\eta$, the average $\left( {\zeta ,\eta } \right)$ pair for different number of antennas is presented in Fig.~\ref{fig:trade-off MISO}. {\rc It is observed that the optimal $\zeta$ grows as number of antennas are increased.}
%%%%%%%%%%%%%%%%%%%%%%%%%%%%%%%%%%%%Conclusion%%%%%%%%%%%%%%%%%%%%%%%%%%%%%%%%%%%%%%%%%%%%%%%%%%%%%%%%
\section{Conclusion}  \label{sec:con}
%%%%%%%%%%%%%%%%%%%%%%%%%%%%%%%%%%%%%%%%%%%%%%%%%%%%%%%%%%%%%%%%%%%%%%%%%%%%%%%%%%%%%%%%%%%%%%%%%%%%%%
In this work, we studied the secrecy energy efficiency, $\zeta$, and its trade-off with the secrecy spectral efficiency, $\eta$, in MISO and SISO wiretap channels. Optimal beamformer
was designed to maximize $\zeta$ for the cases with and without considering the minimum required $\eta$ (i.e., $\eta_{0}$) at the receiver in a power limited system. {\rca We saw that 
as $\eta_{0}$ increases, the performance of the optimal beamformer and the ZF beamformer designs gets closer. Furthermore, as the number of antennas decreases, the performance gap between the optimal and the ZF design increases.} It was observed that there is a specific $\eta$ below which increasing $\eta$ leads to higher secrecy energy efficiency (i.e., $\zeta$), and above which the opposite trend occurs. Depending on the power value corresponding to the optimal $\zeta$, increasing $\eta$ can increase or decrease $\zeta$. In addition, it was shown that adding more antennas to the transmitter side increases $\zeta$ considerably and sustains the optimal $\zeta$ for a longer range of $\eta_0$.

%%%%%%%%%%%%%%%%%%%%%%%%%%%%%%%%%%%%%%%%%%%%%%%%%%%%%%%%%%%%%%%%%%%%%%%%%%%%%%%%%%%%%%%%%%%%%%%%%%

\begin{thebibliography}{10}
\providecommand{\url}[1]{#1}
\csname url@samestyle\endcsname
\providecommand{\newblock}{\relax}
\providecommand{\bibinfo}[2]{#2}
\providecommand{\BIBentrySTDinterwordspacing}{\spaceskip=0pt\relax}
\providecommand{\BIBentryALTinterwordstretchfactor}{4}
\providecommand{\BIBentryALTinterwordspacing}{\spaceskip=\fontdimen2\font plus
\BIBentryALTinterwordstretchfactor\fontdimen3\font minus
  \fontdimen4\font\relax}
\providecommand{\BIBforeignlanguage}[2]{{%
\expandafter\ifx\csname l@#1\endcsname\relax
\typeout{** WARNING: IEEEtran.bst: No hyphenation pattern has been}%
\typeout{** loaded for the language `#1'. Using the pattern for}%
\typeout{** the default language instead.}%
\else
\language=\csname l@#1\endcsname
\fi
#2}}
\providecommand{\BIBdecl}{\relax}
\BIBdecl

\bibitem{sklavos:Cryptography:2007}
N.~Sklavos and X.~Zhang, \emph{\emph{Wireless Security and Cryptography:
  Specifications and Implementations}}.\hskip 1em plus 0.5em minus 0.4em\relax
  Taylor \& Francis, 2007.

\bibitem{Wyner:1975}
A.~D. Wyner, ``The wire-tap channel,'' \emph{Bell Systems Technical Journal},
  vol.~54, no.~8, pp. 1355--1387, Jan. 1975.

\bibitem{Shuguang:2004}
S.~Cui, A.~Goldsmith, and A.~Bahai, ``Energy-efficiency of {MIMO} and
  cooperative {MIMO} techniques in sensor networks,'' \emph{IEEE J. Sel. Areas
  Commun.}, vol.~22, no.~6, pp. 1089--1098, Aug. 2004.

\bibitem{Belmega:2011}
E.-V. Belmega and S.~Lasaulce, ``Energy-efficient precoding for
  multiple-antenna terminals,'' \emph{IEEE Trans. Signal Process.}, vol.~59,
  no.~1, pp. 329--340, Jan. 2011.

\bibitem{Ng:2012}
D.~Ng, E.~Lo, and R.~Schober, ``Energy-efficient resource allocation for secure
  {OFDMA} systems,'' \emph{IEEE Trans. Veh. Technol.}, vol.~61, no.~6, pp.
  2572--2585, Jul. 2012.

\bibitem{Xiaoming:2013}
X.~Chen and L.~Lei, ``Energy-efficient optimization for physical layer security
  in multi-antenna downlink networks with {QoS} guarantee,'' \emph{IEEE Commun.
  Lett.}, vol.~17, no.~4, pp. 637--640, Apr. 2013.

\bibitem{Wang:2013}
L.~Wang, X.~Zhang, X.~Ma, and M.~Song, ``Joint optimization for energy
  consumption and secrecy capacity in wireless cooperative networks,'' in
  \emph{IEEE Wireless Communications and Networking Conference (WCNC)},
  Shanghai, China, Apr. 2013, pp. 941--946.

\bibitem{Jian:2014}
J.~Chen, X.~Chen, T.~Liu, and L.~Lei, ``Energy-efficient power allocation for
  secure communications in large-scale {MIMO} relaying systems,'' in
  \emph{IEEE/CIC International Conference on Communications in China (ICCC)},
  Shanghai, China, Oct. 2014, pp. 385--390.

\bibitem{Zhang:2014}
H.~Zhang, Y.~Huang, S.~Li, and L.~Yang, ``Energy-efficient precoder design for
  {MIMO} wiretap channels,'' \emph{IEEE Commun. Lett.}, vol.~18, no.~9, pp.
  1559--1562, Sep. 2014.

\bibitem{Oggier:2008}
F.~Oggier and B.~Hassibi, ``The {MIMO} wiretap channel,'' in
  \emph{International Symposium on Communications, Control and Signal
  Processing (ISCCSP)}, Malta, Mar. 2008, pp. 213--218.

\bibitem{Barros:2006}
J.~Barros and M.~Rodrigues, ``Secrecy capacity of wireless channels,'' in
  \emph{IEEE International Symposium on Information Theory}, Seattle, WA, Jul.
  2006, pp. 356--360.

\bibitem{Maio:2011}
A.~De~Maio, Y.~Huang, D.~Palomar, S.~Zhang, and A.~Farina, ``Fractional {QCQP}
  with applications in {ML} steering direction estimation for radar
  detection,'' \emph{IEEE Trans. Signal Process.}, vol.~59, no.~1, pp.
  172--185, Jan. 2011.

\bibitem{Wenbao}
{W. Ai, Y. Huang, and S. Zhang}, ``New results on hermitian matrix rank-one
  decomposition,'' \emph{Mathematical Programming}, vol. 128, no. 1-2, pp.
  253--283, Jun. 2011.

\bibitem{horn1990matrix}
R.~Horn and C.~Johnson, \emph{Matrix Analysis}.\hskip 1em plus 0.5em minus
  0.4em\relax Cambridge University Press, 1990.

\bibitem{Siegfried:1983}
S.~Schaible and T.~Ibaraki, ``Fractional programming,'' \emph{European Journal
  of Operational Research}, vol.~12, no.~4, pp. 325--338, Apr. 1983.

\bibitem{Dinkelbach:Dinkelbach}
W.~Dinkelbach, ``On nonlinear fractional programming,'' \emph{Management
  Science}, vol.~13, no.~7, pp. 492--498, Mar. 1967.

\bibitem{3GPP}
\BIBentryALTinterwordspacing
3GPP, ``3rd generation partnership project, technical specification group radio
  access network, coordinated multi-point operation for lte physical layer
  aspects,'' Technical report 36.819, 2011-2012. [Online]. Available:
  \url{http://www.3gpp.org}
\BIBentrySTDinterwordspacing

\end{thebibliography}
\end{document}